\documentclass[10pt]{article}
\usepackage{epsfig}
\begin{document}
\title{\bf Experimental study of the $e^+e^-\to\pi^+\pi^-\pi^0$ reaction by SND
       detector in the energy range ~~$\sqrt[]{s}=0.42$~--~$1.38$ GeV}
\author{M. N. Achasov\thanks{\bf e-mail:achasov@inp.nsk.su}, K.I.Beloborodov,
        A.V.Berdyugin,\\
	A.G.Bogdanchikov, A.V.Bozhenok, A.D.Bukin, D.A.Bukin, \\
	T.V.Dimova, V.P.Druzhinin, V.B.Golubev, A.A.Korol, \\
	S.V.Koshuba, E.V.Pakhtusova, E.A.Perevedentsev, E.E.Pyata, \\
	S.I.Serednyakov, Yu.M.Shatunov, V.A.Sidorov, Z.K.Silagadze, \\
	A.A.Valishev, A.V.Vasiljev\\
	\it \small
        Budker Institute of Nuclear Physics  \\
	\it \small
        Siberian Branch of the Russian Academy of Sciences, \\
        \it \small
        Novosibirsk State University, \\
        \it \small
        Laurentyev 11, Novosibirsk, \\
        \it \small
        630090, Russia \\
        Presented by M.N.Achasov}

\date{}
\maketitle

\begin{abstract}
 The review of the SND results of the $e^+e^-\to\pi^+\pi^-\pi^0$ process study
 in the energy range $\sqrt[]{s}=0.42$ -- $1.38$ GeV at VEPP-2M
 collider, based on about $2\times 10^6$ selected events, is presented.
 The total cross section, parameters of the $\rho$, $\omega$, $\phi$
 resonances, and $\omega^\prime$, $\omega^{\prime\prime}$ states were obtained. 
 It was found that $\rho\pi$ and $\omega\pi^0$ intermediate states describe
 the reaction dynamics. The experimental data cannot be described by a sum of
 only $\omega$, $\phi$, $\omega^\prime$ and $\omega^{\prime\prime}$ resonances
 contributions. This can be interpreted as a manifestation of the 
 $\rho\to 3\pi$ decay, suppressed by $G$-parity, with relative probability
 $B(\rho\to 3\pi) = (1.01\pm^{0.54}_{0.36}\pm 0.034) \times 10^{-4}$.  
 
\end{abstract}

\maketitle

\section{Introduction}
 The $e^+e^-\to \pi^+\pi^-\pi^0$ cross section at low energies is determined
 by the transitions of light vector mesons V
 ($V=\omega,\phi,\omega^\prime,\omega^{\prime\prime}$) into the final state:
 $V\to\rho\pi\to 3\pi$. The mesons with zero isospin have large branching
 ratios: $B(\omega\to 3\pi) \simeq 0.9$, $B(\phi\to 3\pi) \simeq 0.15$
 \cite{pdg}, $B(\omega^\prime\to 3\pi)\sim 1$,
 $B(\omega^{\prime\prime}\to 3\pi)\sim 0.5$ \cite{pi3mhad}.
 The process can also proceed via mechanism suppressed by the G-parity:
 $V\to\omega\pi^0\to 3\pi$ \cite{pi3mhad,thrhoom} or
 $V\to\rho\pi\to 3\pi$ ($V=\rho,\rho^\prime,\rho^{\prime\prime}$).
 The study of the reaction allows to determine the vector
 mesons parameters and provide information on the $OZI$ rule violation in
 the $\phi\to 3\pi$ decay and on the $G$-parity violation in the
 processes $\rho\to 3\pi$.
 The process $e^+e^- \to  3\pi$ in the energy region~ $\sqrt[]{s}$
 below 2.2 GeV was studied in several experiments during the last 30 years
 \cite{vse3pi,dm2,dplkloe}.
 Recently the process  $e^+e^-\to 3\pi$ was also studied by the
 Spherical Neutral Detector (SND)\cite{pi3mhad,sndmhad,phi98,dplphi98,pi3omeg},
 the process dynamics was analyzed and the cross section was measured in the
 energy region $\sqrt[]{s}$ from 420 to 1380 MeV. This talk is a review of the 
 SND results.
\section{Data processing}
 The Spherical Neutral Detector (SND) \cite{snd} has operated since 1995 up to
 2000 at VEPP-2M \cite{vepp2m} $e^+e^-$ collider in the energy range from
 0.36 to 1.38 GeV. During six experimental years SND had collected data with
 integrated luminosity about 30 pb$^{-1}$. 
 During the experimental runs, the first-level trigger \cite{snd} selects
 events with energy deposition in the calorimeter more than 180 MeV and with
 two or more charged particles. For analysis, events containing two charged
 and two or three neutral particles were selected. Extra photons in
 $e^+e^-\to 3\pi$ events can appear because of the overlap with the beam
 background or nuclear interactions of the charged pions
 in the calorimeter. Under these conditions the background sources are
 $e^+e^-\to e^+e^-\gamma\gamma$, $e^+e^-\gamma$,
 $\pi^+\pi^-(\gamma)$, $\mu^+\mu^-(\gamma)$,
 $2\pi^\pm2\pi^0$, $K^+K^-$, $K_SK_L$ processes.
 To reject the collinear background, the cut on 
 $\Delta\phi$ of the charged particles was imposed: $|\Delta\phi|> 5^\circ$.
 To suppress the  $e^+e^-\gamma\gamma$ events an energy deposition of the
 charged particles in the calorimeter was required to be small:
 $E_{cha} < 0.5\cdot \sqrt[]{s}$.
 The cut on dE/dx energy losses in the drift chamber rejected the $K^+K^-$
 events in the vicinity of the $\phi$ meson peak:
 $(dE/dx)<3\cdot(dE/dx)_{min}$. Then a kinematic fit was performed under the
 following constraints: the charged particles are assumed to be pions, the
 system has zero total momentum, the total energy is $\sqrt[]{s}$, and the
 photons originate from the $\pi^0 \to \gamma\gamma$ decays. The cut on the
 $\chi^2_{3\pi}$ was applied: $\chi^2_{3\pi}<20$ at
 $\sqrt{s}<1030$ MeV, and $\chi^2_{3\pi}<5$ at $\sqrt{s}>1030$ MeV.
 In the energy region above 900 MeV for additional suppression of the
 $e^+e^-\to 2\pi^\pm2\pi^0$ and $K_SK_L$ background, the events with exactly
 two photons were selected. The number of selected events is presented in the
 Table~\ref{tab1}.
\begin{table}[t]
\begin{tabular}[b]{cccc}
\hline
$\sqrt[]{s}$& below 980 MeV &  from 980 to 1060 MeV & from 1060 to 1380 MeV \\
\hline
$N_{events}$& $1.2\times 10^6$& $5 \times 10^5$ & $6 \times 10^3$ \\
\hline
\end{tabular}
\caption{The number of selected $e^+e^-\to 3\pi$ events.}
\label{tab1}
\end{table}

\begin{figure}
\includegraphics[width=0.5\textwidth]{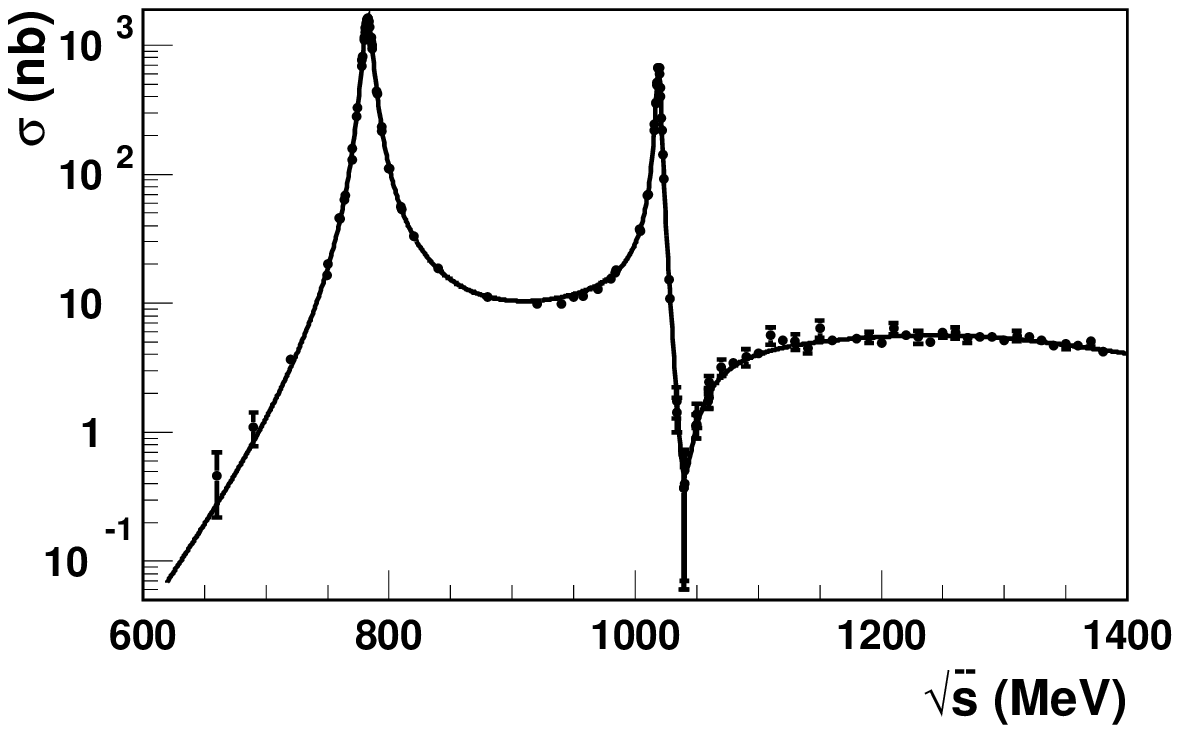}
\hfill
\includegraphics[width=0.5\textwidth]{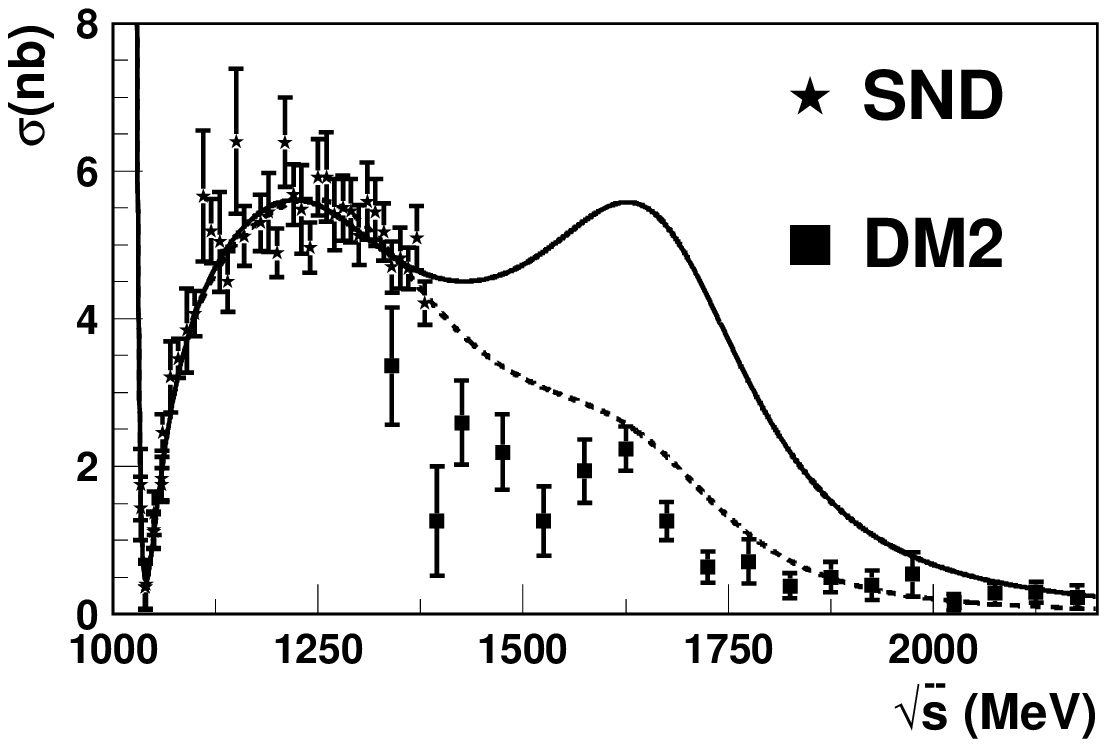}
\\
\vspace{-0.5cm}
\caption{The $e^+e^-\to 3\pi$ cross section. The results of the SND
         \cite{phi98,pi3mhad,pi3omeg} and DM2 \cite{dm2} are shown. Dashed
	 curve corresponds to the fit under assumption that a relative bias
	 between the SND and DM2 data exists (DM2 data were scaled by a factor
	 of 1.5). Solid curve is the result of the fitting to the
         SND data only.}
\label{cstot}
\end{figure}

 The cross section was calculated as the ratio of the number of selected
 events to integrated luminosity, detection efficiency obtained by
 Monte Carlo simulation, and radiative correction for the initial state
 calculated according to Ref.\cite{fadin}.
 The obtained cross section is shown in Fig.\ref{cstot}.
 The total systematic error $\sigma_{sys}\simeq 4$ -- 5\%
 includes the errors of detection efficiency, integrated luminosity, a 
 model error, and an error due to background subtraction. 

\section{Data analysis}
 In analysis of the $e^+e^-\to 3\pi$ reaction we took into account the 
 $\rho\pi$ and $\omega\pi$ transition mechanisms, possible $\rho^\prime\pi$
 transition, and the interaction of the $\rho$ and $\pi$ mesons in the final
 state \cite{akfaz}:
$$
{d\sigma \over dm_0 dm_+} = { {4\pi\alpha} \over {s^{3/2}} }
{{|\vec{p}_+ \times \vec{p}_-|^2} \over {12\pi^2\mbox{~}\sqrt[]{s}}}
m_0m_+ \cdot |F|^2,
$$
$$
|F|^2 = \Biggl| A_{\rho\pi}(s) \cdot \biggl(  \sum_{i=+,0,-}
{ g_{\rho\pi\pi} \over {D_\rho(m_i)Z(m_i)}} + a_{3\pi} \biggr) +
A_{\omega\pi}(s)
{\Pi_{\rho\omega}g_{\rho\pi\pi}\over D_\rho(m_0) D_\omega(m_0)} \Biggr|^2.
$$
 A possible $\rho^\prime\pi$ contribution was written as a constant term,
 because we searched it in the vicinity of the $\phi$ meson
$$
a_{3\pi} =  {A_{\rho^\prime\pi}(s) \over A_{\rho\pi}(s)}  \sum_{i=+,0,-}
{ g_{\rho^\prime\pi\pi} \over {D_{\rho^\prime}(m_i)}},
$$

The $\rho\pi$ and $\omega\pi$ amplitudes are
$$
A_{\rho\pi}(s) = {{1}\over\sqrt[]{4\pi\alpha}}
\sum_{V=\omega,\rho,\phi,\omega^\prime,\omega^{\prime\prime}}
{ 
{\Gamma_V m_V^2 \sqrt[]{m_V\sigma(V \to 3\pi)}} \over 
  {D_V(s)}\sqrt[]{W_{\rho\pi}(m_V) }}
{e^{i\phi_V}}, \mbox{~}
A_{\omega\pi}(s) =
\sum_{V=\rho,\rho^\prime,\rho^{\prime\prime}}
{ {g_{\gamma V}g_{\rho\omega\pi}} \over D_V(s) }.
$$
 From the dipion mass spectra analysis in the $\phi$-meson energy region
 \cite{dplphi98}, it was found that the experimental data can be described
 with $e^+e^-\to\rho\pi\to 3\pi$ transition only. The value of the constant
 term obtained by SND is consistent with zero and differs by 2$\sigma$ from
 KLOE result \cite{dplkloe}. The $\rho$ meson mass and width were measured.
 The mass value agrees with the results obtained in other $e^+e^-$
 experiments. The main results of this analysis are presented in
 Table~\ref{tab2}.
 The analysis of the dipion mass spectra in the energy region above 1.1 GeV
 \cite{pi3mhad} has shown that for their description the
 $e^+e^-\to\omega\pi\to 3\pi$ mechanism is required. The phase between
 $e^+e^-\to\omega\pi$ and $e^+e^-\to\rho\pi$ processes amplitudes was measured.

 The $e^+e^-\to 3\pi$ cross section measured by SND
 \cite{phi98,pi3mhad,pi3omeg} was analyzed together with the DM2 \cite{dm2}
 results on the $e^+e^-\to 3\pi$ and $\omega\pi^+\pi^-$ processes.
 The SND and DM2 measurements agree poorly. So, to take into account possible
 relative systematic shift between experiments, the DM2 cross section was
 multiplied by a factor of 1.5 \cite{pi3mhad,pi3omeg}.
 It was found that for the good description of the data, the
 $\omega$, $\rho$, $\phi$, $\omega^\prime$ and $\omega^{\prime\prime}$
 contributions should be taken into account \cite{pi3omeg}.
 The measured $\omega$ and $\phi$ mesons parameters are shown in
 Table~\ref{tab2} and $\omega^\prime$ and $\omega^{\prime\prime}$ parameters in
 Table~\ref{tab3}.

 The conventional view on the $OZI$ suppressed  $\phi\to 3\pi$ decay
 is that it proceeds through $\phi$-$\omega$ mixing, i.e.
 in the wave function of the $\phi$-meson which is dominated by $s$ quarks,
 there is an admixture of $u$ and $d$ quarks.
 An alternative to the $\phi$-$\omega$ mixing is the direct decay
 \cite{vsephiom}.
 Analysis of the $\Gamma(\phi\to e^+e^-)/\Gamma(\omega\to e^+e^-)$ ratio
 and $g_{\phi\rho\pi}$ and $g_{\omega\rho\pi}$ coupling constants obtained in
 SND experiments indicates that the direct transition is preferable to the
 $\phi$-$\omega$ mixing as the main mechanism of the $\phi\to 3\pi$
 decay \cite{pi3omeg}.
    
 The $\omega^\prime$ and $\omega^{\prime\prime}$ parameters obtained from the
 fits (Table~\ref{tab3}) should be considered as rather approximate estimation
 of the $\omega^\prime$ and $\omega^{\prime\prime}$ resonances main parameters.
 To measure the parameters of these states precisely new data above 
 1.4 GeV required.
 
\begin{table}[t]
\begin{tabular}[b]{llll}
\hline
      & SND & Other data \\
\hline
$m_\rho$, MeV& 775.0 $\pm$ 1.3& 775.9 $\pm$ 0.5&(PDG-2002) \\
$\Gamma_\rho$, MeV& 150.4 $\pm$ 3.0& 147.9 $\pm$ 1.3&(PDG-2002) \\
$m_{\rho^\pm}-m_{\rho^0}$, MeV& -1.3 $\pm$ 2.3&0.4 $\pm$ 0.7 $\pm$ 0.06&(KLOE\cite{dplkloe}) \\
$|a_{3\pi}| \times 10^5$, MeV$^{-2}$& 0.01 $\pm$ 0.34&0.7 $\pm$ 0.1&(KLOE\cite{dplkloe}) \\
\hline
$m_\omega$, MeV&782.79 $\pm$ 0.08 $\pm$ 0.09&782.57 $\pm$ 0.12&(PDG-2002)\\
$\Gamma_\omega$, MeV&8.68 $\pm$ 0.04 $\pm$ 0.24&8.44 $\pm$ 0.09&(PDG-2002)\\
$\sigma(\omega\to 3\pi)$, nb&1615 $\pm$ 9 $\pm$ 57&636 $\pm$ 27&(PDG-2002)\\
\hline
$\sigma(\phi\to 3\pi)$, nb&657 $\pm$ 10 $\pm$ 37&1484 $\pm$ 29&(PDG-2000) \\
$\phi_{\phi}$, degree&163 $\pm$ 3 $\pm$ 6&158 --- 172& \cite{faza} \\
\hline
\end{tabular}
\caption{The results of the dipion mass spectra analysis \cite{dplphi98},
         results of $\omega\to 3\pi$ and $\phi\to 3\pi$ decays study 
	 \cite{pi3omeg,phi98}.}
\label{tab2}
\end{table}

 It was found that the experimental data cannot be described by a sum of
 $\omega$, $\phi$, $\omega^\prime$ and $\omega^{\prime\prime}$ resonances
 contributions. This can be interpreted as a manifestation of the
 $\rho\to 3\pi$ decay suppressed by $G$-parity. The obtained parameters of the
 decay $B(\rho\to 3\pi)=(1.01\pm^{0.54}_{0.36}\pm 0.34)\times 10^{-4}$ and
 $\phi_{\rho}=-135\pm^{17}_{13} \pm 9$ degree are in
 agreement with the theoretical values expected from the $\rho$-$\omega$
 mixing $B(\rho\to 3\pi)=(0.4$ -- $0.6)\times 10^{-4}$ and
 $\phi_{\rho}\simeq -90$ degree.

 Using the $e^+e^-\to 3\pi$ cross section obtained by SND
 detector, the contribution to the anomalous magnetic moment of the muon due
 to the $\pi^+\pi^-\pi^0$ intermediate state was calculated
 $a_\mu(3\pi,\sqrt[]{s}<1.38\mbox{GeV})=(458 \pm 2 \pm 17) \times 10^{-11}.$
\begin{table}[t]
\begin{center}
\begin{tabular}{lll}
\hline
  $V$&$\omega^\prime$& $\omega^{\prime\prime}$ \\ \hline
  $m_V$, MeV&1400 $\pm$ 50 $\pm$ 130&1770 $\pm$ 50 $\pm$ 60 \\
   $\Gamma_V$, MeV&870 $\pm^{500}_{300}\pm$ 450&490 $\pm^{200}_{150}\pm$ 130 \\
   $\sigma(V\to 3\pi)$, nb&4.9 $\pm$ 1.0 $\pm$ 1.6&5.4 $\pm^{0.2}_{0.4}\pm$ 3.9 \\
   $\sigma(V\to \omega\pi^+\pi^-)$, nb&&1.9 $\pm$ 0.4 $\pm$ 0.6 \\
   $B(V\to e^+e^-)$&$\sim$6.5$\times 10^{-7}$&$\sim$ 1.6 $\times 10^{-6}$ \\
   $\Gamma(V\to e^+e^-)$, eV&$\sim$ 570& $\sim$ 860\\
   $B(V\to 3\pi)$&$\sim$1&$\sim$ 0.65 \\
   $B(V\to \omega 2\pi)$&&$\sim$ 0.35 \\
\hline
\end{tabular}
\caption{The $\omega^\prime$ and $\omega^{\prime\prime}$ parameters obtained
         from the fit \cite{pi3omeg,pi3mhad} of SND and DM2 data.}
\label{tab3}
\end{center}
\end{table}

\section{Conclusion}
 The $e^+e^-\to  3\pi$ cross section was measured in the SND
 experiment at the VEPP-2M collider in the energy region~ $\sqrt[]{s}$ below
 1380 MeV. The experimental data were analyzed in the framework of the
 generalized vector meson dominance model. It was found that the $\omega\pi$
 and $\rho\pi$ intermediate states describe the process dynamics.
 The $\omega$ and $\phi$ mesons parameters were obtained and
 parameters of the $\omega^\prime$, $\omega^{\prime\prime}$ resonances
 were estimated. Experimental data cannot be described
 by a sum of $\omega$, $\phi$, $\omega^\prime$ and $\omega^{\prime\prime}$ 
 contributions. This can be interpreted as a manifestation of the
 $\rho\to 3\pi$ decay. The SND study of the $e^+e^-\to 3\pi$ process was
 reported in Ref.\cite{phi98,dplphi98,pi3mhad,pi3omeg}.
 Now the VEPP-2000 collider with the maximum center-of-mass energy  2 GeV is
 under construction \cite{vepp2000}.
 The $e^+e^-\to 3\pi$ process study will be continued in future experiments
 with SND detector at VEPP-2000.

\section{Acknowledgments}
 
 This work was supported in part by Presidential Grant 1335.2003.2 for
 support of Leading Scientific Schools and by Russian Science Support
 Foundation.

\end{document}